\documentstyle[12pt,epsfig]{article}

\newlength{\dinwidth}
\newlength{\dinmargin}
\def\lapproxeq{\lower .7ex\hbox{$\;\stackrel{\textstyle
<}{\sim}\;$}}
\def\gapproxeq{\lower .7ex\hbox{$\;\stackrel{\textstyle
>}{\sim}\;$}}

\def\beq{\begin{equation}}
\def\eeq{\end{equation}}
\def\bea{\begin{eqnarray}}
\def\eea{\end{eqnarray}}

\def\GeV{\rm GeV}

\begin{document}
\titlepage
\begin{flushright}
IPPP/06/13 \\
DCPT/06/26 \\
Cavendish-HEP-2006/07 \\

\end{flushright}

\vspace*{0.5cm}

\begin{center}
{\Large \bf MRST partons generated in a fixed-flavour scheme }

\vspace*{1cm}
\textsc{A.D. Martin$^a$, W.J. Stirling$^a$
and R.S. Thorne$^{c,}$\footnote{Royal Society University Research Fellow.}} \\

\vspace*{0.5cm} $^a$ Institute for Particle Physics Phenomenology,
University of Durham, DH1 3LE, UK \\
$^c$ Cavendish Laboratory, University of Cambridge, \\ JJ Thomson Avenue,
Cambridge, CB3 0HE, UK
\end{center}

\vspace*{0.5cm}

\begin{abstract}
We generate fixed three- and four-light-flavour sets of partons using 
MRST2004 partons as input.
We show that it is important to set $n_f=3$ in the strong coupling, as 
well as in the
splitting and coefficient functions, in order to obtain a consistent set 
of fixed-flavour partons.
We compare the description of data using partons in both variable- and 
fixed-flavour-number-schemes.
\end{abstract}

\vspace*{0.5cm}

In deep inelastic scattering, and ``hard'' proton-proton 
(or antiproton-proton) high-energy collisions, the
scattering proceeds via the partonic constituents of the hadron.  To predict 
the rates of the various
processes a set of universal parton distribution functions is required.  
These quark and gluon
distributions are usually given in a
variable-flavour-number-scheme (VFNS) in which the number of active quark 
flavours increases from
$n_f=3$ to $n_f=4$, and then to $n_f=5$, as $Q^2$ increases above about 
$m_c^2$, and then above
about $m_b^2$.  Here $Q^2$ is the magnitude of the square of the momentum 
transfer in the process.

However there are practical reasons for also generating a set of up-to-date 
partons in the fixed three-light-flavour
($n_f=3$) and fixed four-light-flavour
($n_f=4$) schemes.  For many exclusive or semi-inclusive processes the 
theoretical predictions for the hard scattering cross sections are, as yet, 
only available in the fixed-flavour-number-scheme (FFNS)
where all the heavy flavours are produced in the hard cross-section.
Thus fixed-flavour partons 
are needed in existing Monte Carlos, see, for 
example, \cite{harris,mangano,mcatnlo}.  Also, despite the availability of a 
number of general mass variable-flavour number schemes, extractions of 
partons from fits to structure function data are often performed using a 
FFNS, so it is useful to have available an 
up-to-date parton set in order to make consistent comparisons and to judge the real 
differences between alternative parton sets obtained using the same theoretical 
framework.   

Here we generate a set of FFNS partons by evolving up in $Q^2$ from a set of 
partons at the input scale, $Q^2_0=1~{\rm GeV}^2$, obtained in a VFNS global analysis of 
deep inelastic and related hard scattering data. The evolution differs from
that used in the VFNS fit to the data in that
\begin{itemize}
\item we never turn on the heavy flavour $(c,b)$ contributions,
\item we set $n_f=3$ in the splitting and coefficient functions,
\item we keep $n_f=3$ in the coupling $\alpha_s(Q^2)$.
\end{itemize}
With respect to the last point, we usually define the coupling via the 
$n_f=4$ definition of $\Lambda_{\rm QCD}$;
for example for the (${\overline {\rm MS}}$, NLO) MRST2004 set of partons \cite{MRST04} we have 
$\Lambda^{(n_f=4)}_{\rm QCD}=347~{\rm MeV}$.  We can convert this to a
$n_f=3$ definition of $\Lambda_{\rm QCD}$ using the NLO relation \cite{Marciano}
\beq
\Lambda^{(3)}~=~\Lambda^{(4)}~\left(
\frac{m_c}{\Lambda^{(4)}}\right)^{\frac{2}{27}}
\left[ 2~{\rm ln}\left( \frac{m_c}{\Lambda^{(4)}}    
\right) \right]^{\frac{107}{2025}}.
\eeq
This relationship guarantees that the two couplings $\alpha_S$ are identical 
below $m_c^2$. However the
three-flavour coupling runs more quickly above the charm quark transition 
point, and hence becomes
smaller.  The discrepancy between the speed of evolution increases further 
at the bottom transition point,
and the three-flavour coupling has a value $\alpha^{(3)}_S(M_Z^2)=0.107$ 
compared to the value 
$\alpha^{(5)}_S(M_Z^2)=0.120$ obtained in the VFNS fit \cite{MRST04}.  
A similar behaviour is found for
the CTEQ5 FFNS partons \cite{CTEQ5}, where, there, the three-flavour coupling 
has a value $\alpha^{(3)}_S(M_Z^2)=0.106$ compared to the value 
$\alpha^{(5)}_S(M_Z^2)=0.118$.  Thus, at high $Q^2$, 
$\alpha^{(3)}_S$ and $\alpha^{(5)}_S$ are
significantly different quantities. Forcing $\alpha^{(3)}_S$ to take the 
value $\alpha^{(3)}_S(M_Z^2)=0.120$
would result in an enhanced coupling for $Q^2 \sim 10~{\rm GeV}^2$, which 
would be inappropriate for describing data at these scales.

We emphasise that it is important to do the FF evolution consistently. 
We note that the determination of FFNS partons is very frequently done 
incorrectly, with
a ``variable-flavour'' strong coupling, $\alpha_S$, used along 
with coefficient and splitting functions renormalized in 
a $n_f=3$ FFNS.  Indeed this inconsistency between the renormalization 
scheme used for the coefficient functions and
for the coupling was apparent in the original calculation \cite{laenen} of 
the FFNS coefficient functions for
deep inelastic scattering.  The authors\footnote{We thank Eric Laenen for 
clarifying this point.} were fully
aware of this discrepancy, but used both a VFNS coupling and VFNS partons 
since only VFNS partons were then available and because the error induced was 
small compared with other uncertainties. It is no longer the case that the 
error made from the wrong choice of coupling is so unimportant.

\begin{figure}
\begin{center}
\centerline{\epsfxsize=0.7\textwidth\epsfbox{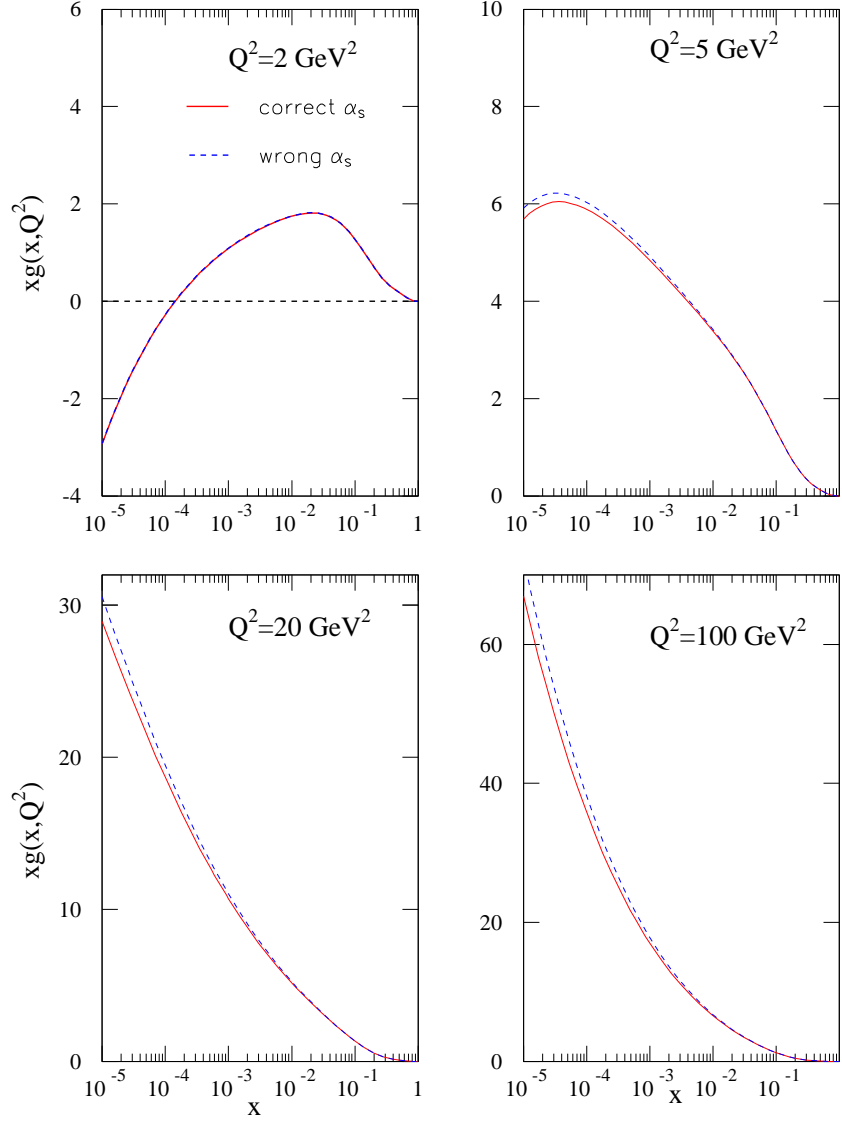}}
\vspace{0.3cm}
\caption{The effect on the gluon distribution of, using in the 3-flavour 
FF evolution, a 
different number of active quark flavours in the coupling and the
structure functions. The wrong $\alpha_S$ corresponds to increasing $n_f$ 
in the coupling at each heavy quark threshold.}
\label{fig:gluwffns}
\end{center}
\end{figure}

An illustration of the effect of such an inconsistency on a physical quantity 
is provided by the evolution
of $F_L$. At leading order the gluon contribution to $F_L$ is
\beq
F_L~=~\alpha_S~C^1_{Lg} \otimes g,
\eeq
and therefore at this order
\beq
\frac{\partial F_L}{\partial {\rm ln}Q^2}~=~-\beta_0 ~\alpha^2_S~ C^1_{Lg} 
\otimes g~+~
\alpha^2_S~ C^1_{Lg} \otimes P^{(0)}_{gg} \otimes g~~+ ~{\rm quark ~term}.
\eeq
Here $\beta_0=(11-\frac{2}{3}n_f)/4\pi$, whereas $P^{(0)}_{gg}$ contains a term
$-(\frac{2}{3}n_f/4\pi)\delta (1-z)$, i.e. the gluon loses momentum to the 
quarks it radiates.  Hence, for example, in going from the $n_f=3$ renormalization scheme 
to the
$n_f=4$ renormalization scheme, the change in these two terms cancels out, 
leaving the physical quantity
independent of choice of number of active quarks $n_f$.  However if the 
coupling and the structure functions use a different number of active quarks, 
as is often done, the cancellation does not occur.  The effect of this 
inconsistent choice on 
the gluon distribution is shown in Fig.\ref{fig:gluwffns}. The gluon evolved 
using the incorrect coupling evolves more quickly than that with the correct 
coupling since the coupling is always larger in the former case. By 
$Q^2=100~\GeV^2$ the incorrect gluon is about $10\%$ larger at $x=0.00001$. 
At $x\sim 0.05$, where there is little change with $Q^2$, the gluons are very 
similar, but at high $x$ the gluon with the incorrect coupling decreases more 
quickly and at $Q^2=100~\GeV^2$ is up to $10\%$ smaller at $x=0.5$. The error 
on the gluon due to the mistake in the choice of coupling is generally of 
similar size 
to the uncertainty due to experimental errors in \cite{MRSTerror}. As such, 
this effect is far from insignificant. A similar effect is seen in the quark
distributions, i.e. evolution is too quick, and some effects due to internal
quark loops are double-counted.   

\begin{figure}
\begin{center}
\centerline{\epsfxsize=0.7\textwidth\epsfbox{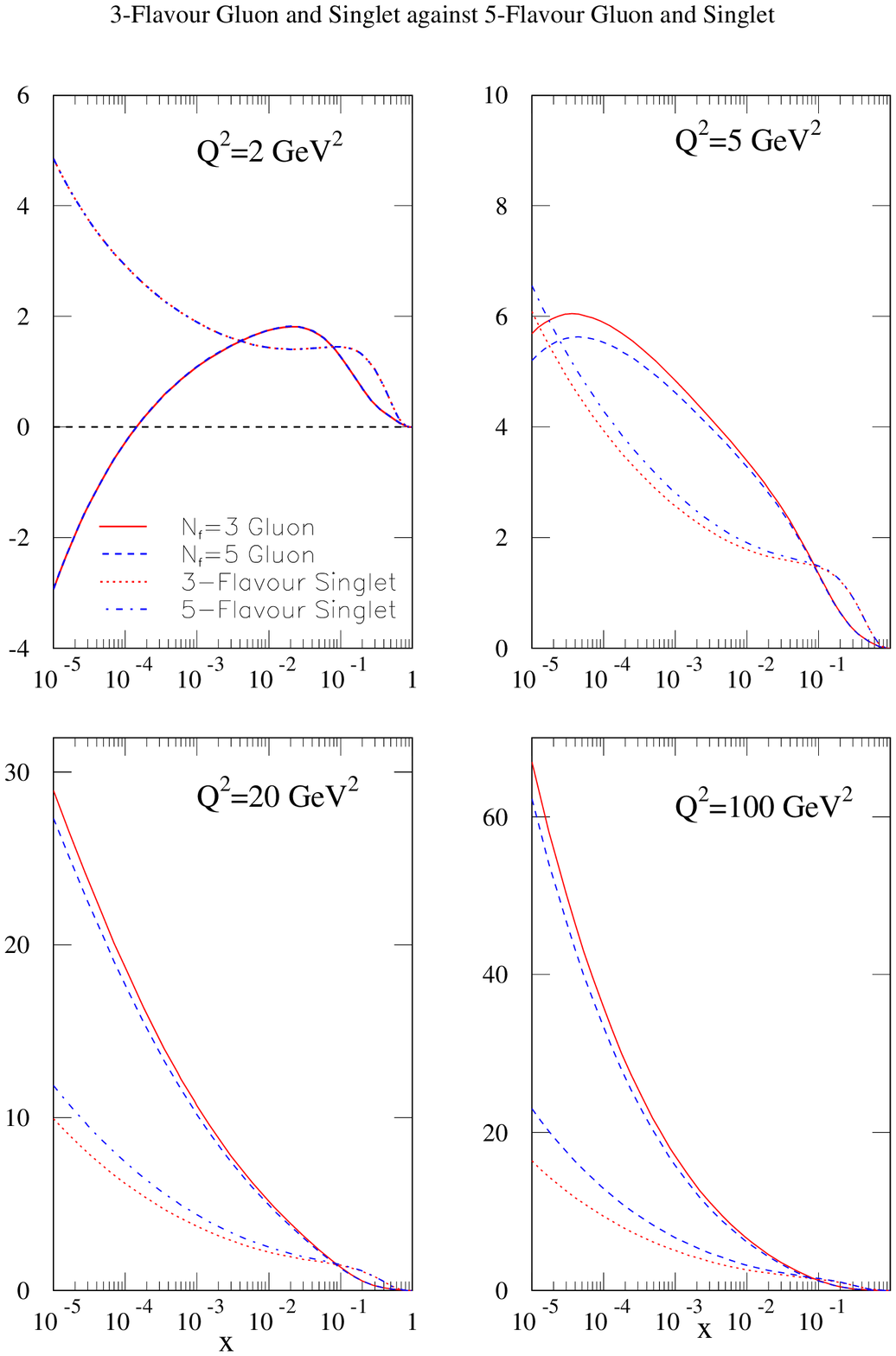}}
\vspace{0.3cm}
\caption{The comparison of the gluon distributions (and the total singlet-quark distributions) 
in the 3-flavour FFNS and the VFNS.}
\label{fig:glusing}
\end{center}
\end{figure}

\begin{figure}
\begin{center}
\centerline{\epsfxsize=0.7\textwidth\epsfbox{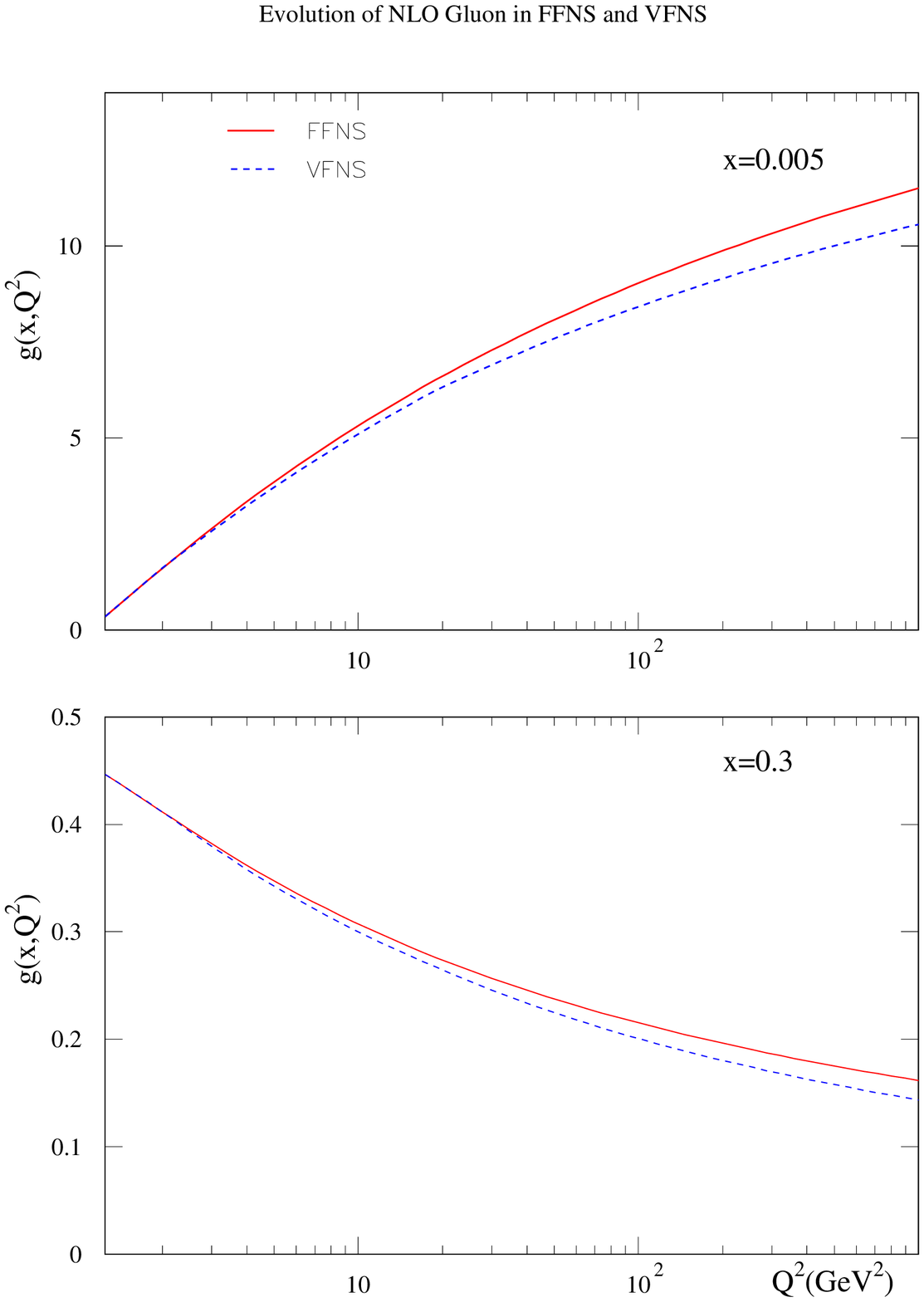}}
\vspace{0.3cm}
\caption{The evolution of the gluon distribution in the 3-flavour 
FFNS and the VFNS, for typical low and high values of $x$.}
\label{fig:gluffvfevol}
\end{center}
\end{figure}

\begin{figure}
\begin{center}
\centerline{\epsfxsize=0.7\textwidth\epsfbox{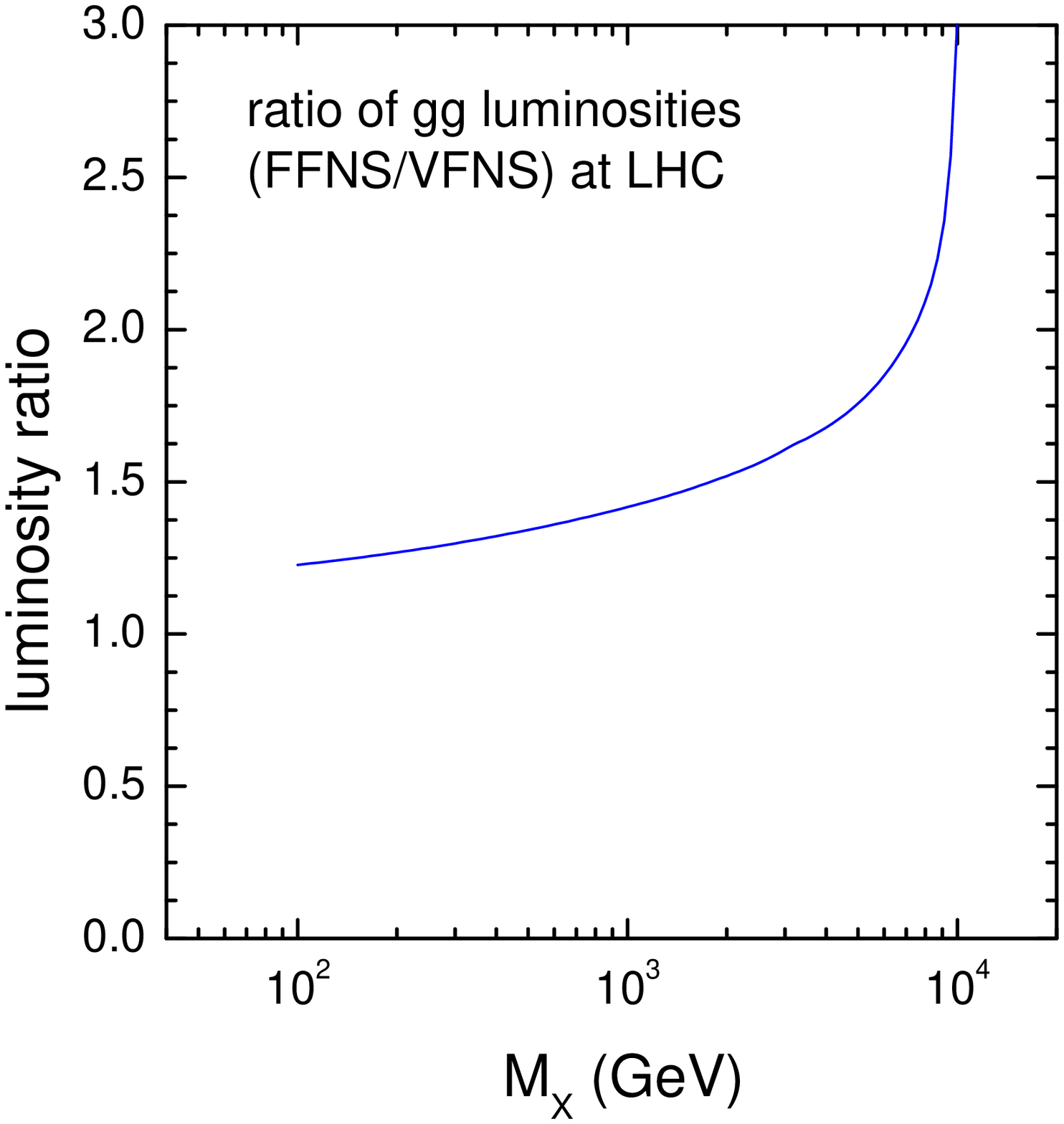}}
\vspace{-3.3cm}
\caption{The effect on the luminosity distribution for $gg \to X$ of using 
FFNS or VFNS partons.}
\label{fig:gglumiff}
\end{center}
\end{figure}

We make the correctly generated 3- and 4-flavour FFNS parton distributions, which we denote by MRST2004FF3 and MRST2004FF4, 
generally available\footnote{These parton sets can be
found at  http://durpdg.dur.ac.uk/hepdata/mrs.html.}. We do this 
at both NLO and also at LO since LO partons are sometimes needed for LO Monte 
Carlo programmes (see e.g. \cite{Herwig}). 
The LO partons are the companions to those in \cite{MRSTNNLO},
i.e. the MRST2001 LO partons, and as in this previous paper we point out that
the LO partons are not able to give a genuinely good global fit and should 
be used with caution. The NLO 3-flavour gluon and
singlet-quark distributions are compared to their VFNS counterparts in Fig.\ref{fig:glusing}. The gluon 
distribution is always bigger in the 3-flavour scheme because of the extra 
radiation of gluons to charm and bottom quarks in the VFNS. This extra 
positive evolution outweighs the decreased evolution due to the lower 
3-flavour coupling at low $x$. At high $x$ both the lower coupling and the 
lack of radiation to quarks slow the decrease of the 3-flavour gluon relative
to that of the VFNS gluon. These two effects are shown in  
Fig.\ref{fig:gluffvfevol}. The 3-flavour gluon is hence always bigger than 
the VFNS gluon for $Q^2> m_c^2$. Conversely, as seen in 
Fig.\ref{fig:glusing}, the singlet quark distribution is always correspondingly
bigger. Almost all of this is difference is due to the presence of the 
charm and bottom quarks, which carry the extra momentum lost by the VFNS 
gluon. In order to highlight the difference between the 3-flavour and 
VFNS gluon we also plot in Fig.\ref{fig:gglumiff} the 
ratio of the corresponding $gg$ luminosity functions
$(\partial {\cal L}^{gg}/\partial \tau,~\tau=M_X^2/s)$ for producing a heavy 
particle of mass $M_X$ at the LHC. The larger 3-flavour gluon always
leads to a larger luminosity, particularly when probing very high $x$ where
the gluons differ most. This highlights the extreme care which must be taken 
in using the correctly defined gluon together with the correctly defined 
coupling and hard cross-section in any predictions for particle production 
at the LHC or Tevatron.    

This latter point also illustrates the difficulty in performing a correct 
global fit to parton distributions with a FFNS, i.e. we need 
exact versions of the NLO hard cross-sections which should be used with 
the FFNS partons. All will have logarithms in $Q^2/m_H^2$, which compensate 
for the missing heavy quark parton distributions, along with other mass 
effects. (Here the subscript $H=c$ or $b$.) However, these are not known 
for all processes. Hence, our 
decision to generate instead sets of FFNS partons which are consistent with
our standard partons below $Q^2=m_c^2$. We are easily able to 
calculate the structure function $F_2(x,Q^2)$ in the FFNS scheme however.
Note that in this scheme the coefficient functions for heavy flavour 
production begin at order $\alpha_S$, so LO corresponds to 
${\cal O}(\alpha_S)$ coefficient functions, NLO to ${\cal O}(\alpha^2_S)$,
${\cal O}(\alpha_S^3)$, {\it etc.}. This is consistent since the 
${\cal O}(\alpha_S)$ coefficient function contains information on the LO
splitting function $P^{(0)}_{qg}$ the ${\cal O}(\alpha^2_S)$ coefficient function
on the NLO splitting function $P^{(1)}_{qg}$, {\it etc.}, so at each order the 
evolution of the heavy flavour structure function is roughly correct.    
The NLO cross-sections for the heavy flavour contributions have  
long existed as semi-analytic code \cite{code} where
the dominant contributions for 
$W^2 \to \infty$ and $W^2 \to 4m_H^2$ (from above) are 
analytic and the rest numerical.  The NNLO coefficient functions for 
heavy flavour production are not known yet, though 
approximate expressions for low $Q^2$ are derived in \cite{nnlovfns}, 
which enables one to define a NNLO
VFNS. As another by-product of the definition of the scheme in 
\cite{nnlovfns} there exist 
faster analytic expressions for the NLO coefficient functions
which are exact for $Q^2/m_H^2 \to \infty$ 
and in some cases for $W^2 \to 4m_H^2$, and 
the $(m_H^2/Q^2)$ remainders are provided by fitting the values to 
analytic functions with a number of free parameters. These final expressions
are slightly approximate,
but the error in $F_2^H(x,Q^2)$ is only $\sim 1\%$, even in the most extreme 
cases.\footnote{This code can also be
found at  http://durpdg.dur.ac.uk/hepdata/mrs.html.}   

\begin{figure}
\begin{center}
\centerline{\hspace{-1.0cm}\epsfxsize=0.45\textwidth\epsfbox{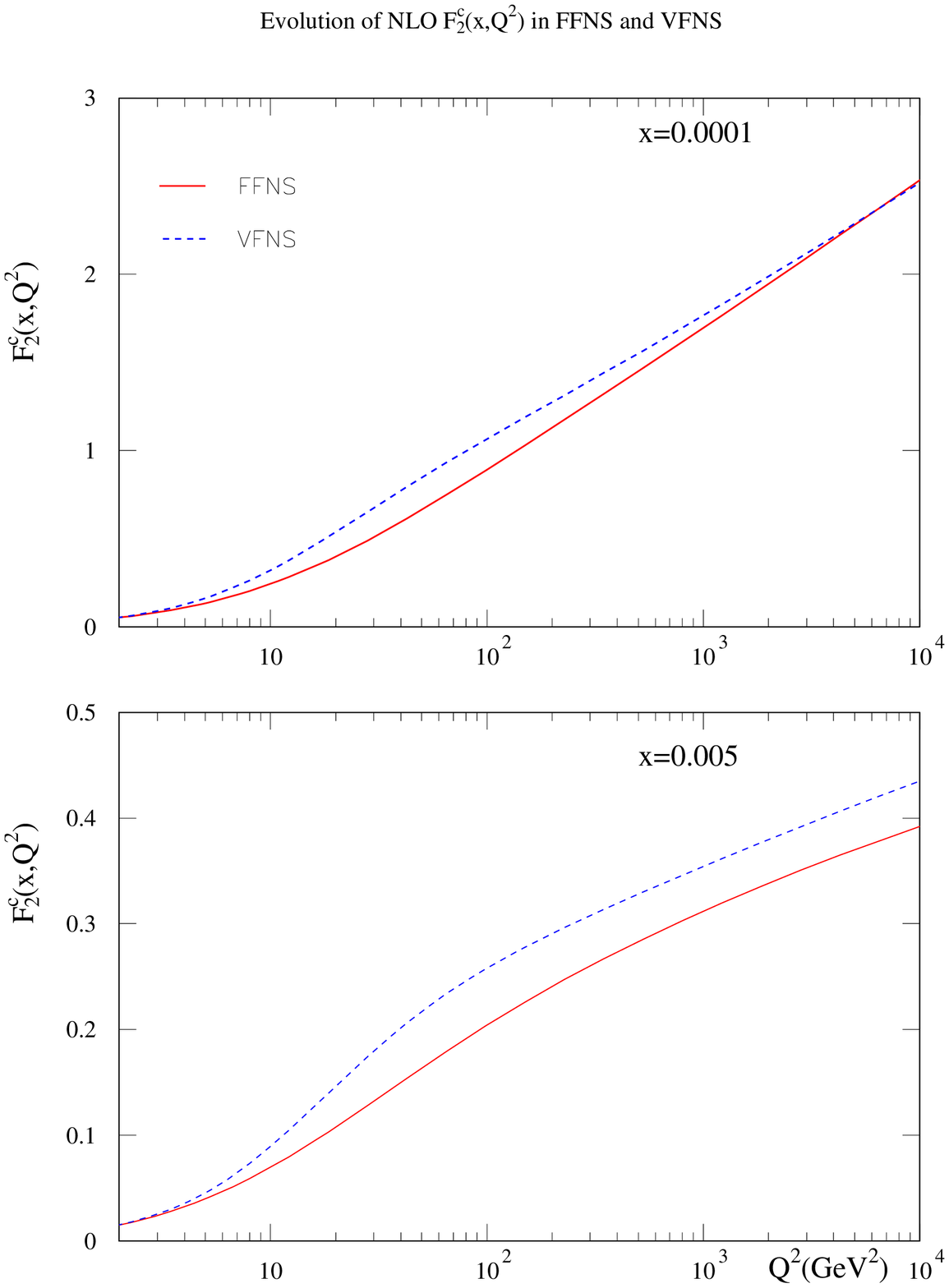}
\hspace{0.9cm}\epsfxsize=0.45\textwidth\epsfbox{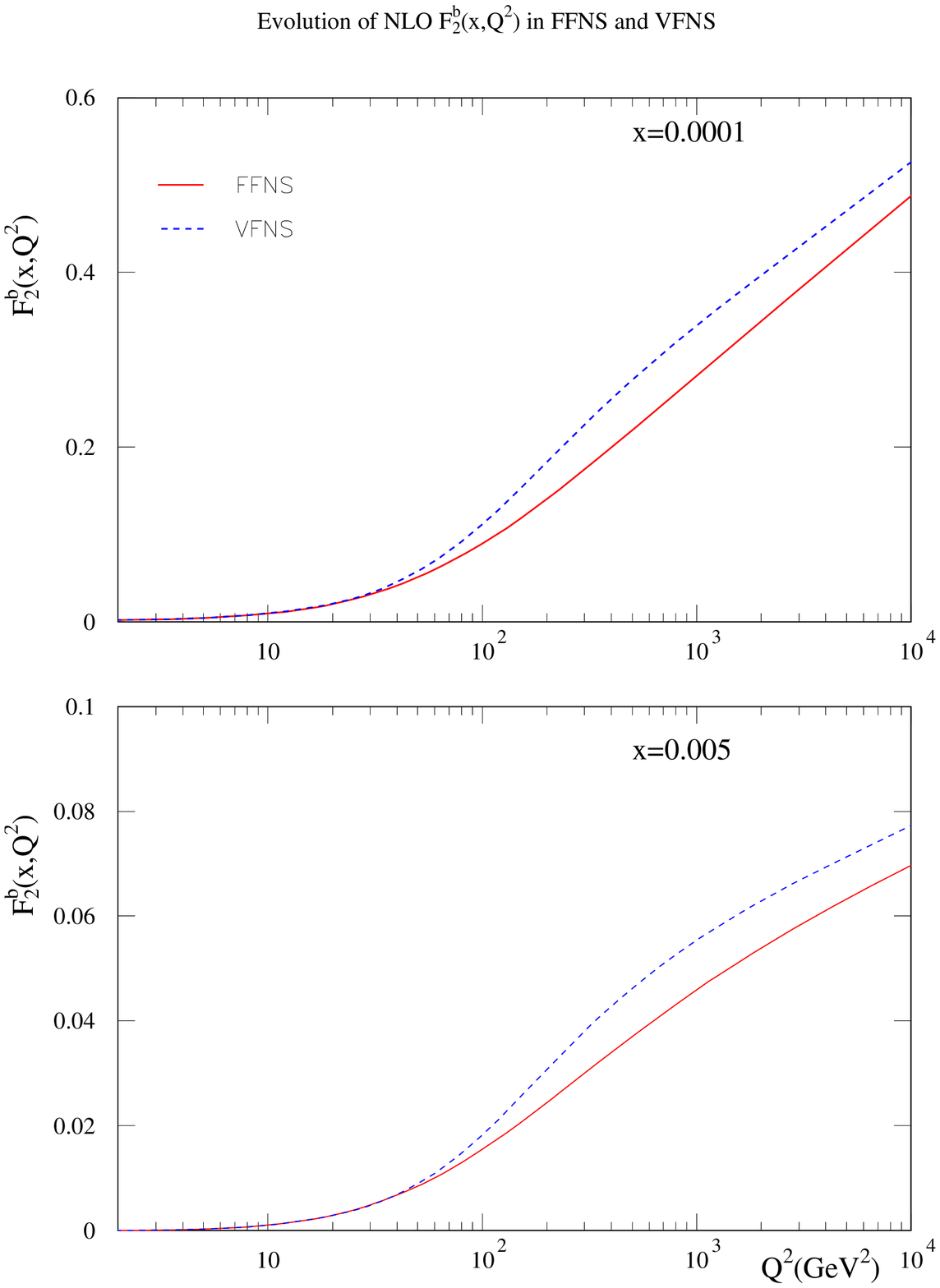}}
\vspace{0.3cm}
\caption{The evolution of $F_2^c(x,Q^2)$ in the 3-flavour FFNS 
and VFNS (left). The evolution of $F_2^b(x,Q^2)$ in the 3-flavour FFNS 
and VFNS (right).}
\label{fig:bottevol}
\end{center}
\end{figure}

The results for the charm structure function and bottom structure function 
in the FFNS compared to that from 
the MRST2004 partons using VFNS coefficient functions (using the scheme of
\cite{trvfns} appropriate for these partons) are shown in 
Fig.\ref{fig:bottevol}, where the factorization scale $\mu^2=Q^2$ has been 
used in both cases. Clearly heavy flavour production is generally suppressed 
in the FFNS compared to the
VFNS.\footnote{This trend has been observed when comparing to recent heavy 
flavour data using a wide variety of partons and scales \cite{Thompson}. 
In previous studies the incorrect
use of the VFNS coupling in fixed-flavour number schemes has often somewhat 
masked the suppression.} However, it is interesting to note that at extremely 
high $Q^2$ the charm structure function in the VFNS for very small $x$ has 
come back towards that in the FFNS and even goes a little below the FFNS 
result. This is because for these very large values of $Q^2/m_c^2$
the extra powers of $\ln Q^2/m_c^2$ resummed by the VFNS procedure are 
accompanied by multiple convolutions of splitting functions containing 
information on the loss of charm quarks due to radiation to gluons and 
even smaller $x$ quarks. This information is missing in the FFNS which only 
contains the first two powers of $\ln Q^2/m_c^2$ and the accompanying 
splitting functions.    
\begin{table}[htb]
\caption{Quality of the FFNS comparison to some data sets, together
with the corresponding $\chi^2$ values for the standard NLO VFNS fit to
the same data. Here the FFNS partons are not obtained by fitting to
data, but rather are generated by a FF ($n_f=3$) evolution from the
VFNS input distributions.}
\begin{center}
\begin{tabular}{|lccc|} \hline
Data set & No.\ of & VFNS & FFNS \\
& data pts & & \\ \hline
 H1 $ep$ & 417 & 427 & 1117 \\
 ZEUS $ep$ & 356 & 279 & 893 \\
 BCDMS $\mu p$ & 179 & 190 & 174 \\
 BCDMS $\mu d$ & 155 & 216 & 242 \\
 NMC $\mu p$ & 126 & 136 & 164 \\
 NMC $\mu d$ & 126 & 103 & 128 \\
 SLAC $ep$ & 53 & 50 & 74 \\
 SLAC $ed$ & 54 & 56 & 58 \\
 E665 $\mu p$ & 53 & 50 & 49\\
 E665 $\mu d$ & 53 & 61 & 63 \\
\hline
\end{tabular}
\end{center}
\end{table}

This general suppression of heavy flavour in 
FFNS means the quality of the match 
to the 
data using the MRST FFNS partons is not expected to be very good in regions 
where the charm contribution to the structure function is important. 
In the Table we show the contributions to $\chi^2$ from various subsets of 
the {\it data fitted} in the global
analysis which yields the NLO MRST2004 (VFNS) partons \cite{MRST04}.  The 
Table compares these values with the
$\chi^2$ obtained from the FFNS partons {\it generated} using the MRST2004 
partons as input at $Q_0^2=1~{\rm GeV}^2$.
We emphasise that the latter partons are not obtained in a new fit to the 
data.  We see that the $\chi^2$ for the
HERA (H1, ZEUS) data is indeed significantly larger using the FFNS partons. 
This is due to the fact that at small $x,~
(\lapproxeq 10^{-2}),~F_2^c$ increases with $Q^2$ up to $30\%$ more slowly 
in the FFNS.  Since $F_2^c$ is a
major component of the total $F_2$ at small $x$, this leads to a 
general undershooting of these HERA data.
Note that the description of the fixed-target (BCDMS, SLAC) structure 
function data is not changed
significantly in going from VFNS to FFNS, while the NMC data is slightly 
sensitive to the charm structure function and is affected a little more.  
This consistency in the fixed-target data is because the slowing down of 
the evolution above $Q^2 \sim m^2_c$,
due to the reduced coupling, is largely compensated by the $n_f$ dependence 
of the NLO splitting functions.
If we were to use the VFNS coupling for the FFNS partons the description of 
the fixed-target data would be worse.
As already noted, the smaller $\alpha_S$ for the FFNS partons at high $Q^2$ is 
accompanied by a larger gluon distribution at
large $x$ due to the slower evolution. These two effects compensate in the 
description of the Tevatron inclusive jet data which, as far as one can tell
without FFNS jet coefficient functions, is similar for both schemes.

We notice that it is the charm (and bottom) data themselves, and hence 
the description of the small $x$ structure function data, that are the most 
sensitive to the choice of the scheme for the heavy flavours, which is,
of course, not unexpected. Indeed, if we do attempt a fit to as much data in 
the global fit as possible in the FFNS scheme\footnote{Only the neutral 
current structure function data may be included completely rigorously. The NLO
FFNS coefficient functions are not calculated for charged current structure 
functions. NLO FFNS coefficient functions are calculated for the Drell-Yan
cross-section \cite{DYhf} but not in sufficiently differential form.} 
our comparison to HERA data is 
always 80 units of $\chi^2$ or so worse than in the VFNS scheme, even 
though the value of the 3-flavour coupling increases significantly in an 
attempt to fit 
the small-$x$ data more successfully. This increase in the coupling is such 
that it causes the fit to fixed-target data to deteriorate. We believe that 
this difficulty in fitting small-$x$ structure function data in a FFNS scheme
would be generally observed if the correct definition of the coupling were 
used and constraints from data sets other than DIS data were applied in such 
fits. Hence, we conclude that it is always preferable, if possible, to 
work in a variable-flavour number scheme for parton distributions. However,
recognising that this is not always possible for current applications, and 
also that there is a general interest in FFNS partons, we use this paper 
to advertise and make available our 3-flavour and 4-flavour scheme partons
generated from the same input as our 2004 VFNS partons, but consistently 
evolved in these two alternative schemes.

\section*{Acknowledgements}

RST would like to thank Bryan Webber for discussions on the use of partons 
in heavy-flavour Monte Carlos and 
Paul Thompson for asking questions which led to the 
investigation of the use of the coupling constant in fixed-flavour schemes
and for comments on the paper. RST thanks
the Royal Society for the award of a University Research Fellowship and ADM 
thanks the Leverhulme Trust for the award of an Emeritus
Fellowship. The IPPP gratefully acknowledges financial support from the UK
Particle Physics and Astronomy Research Council.\\

\vspace{-0.6cm}


\end{document}